\documentclass[journal=apchd5,manuscript=article]{achemso}

\usepackage[version=3]{mhchem} % Formula subscripts using \ce{}
\usepackage{graphicx}% Include figure files
\usepackage{dcolumn}% Align table columns on decimal point
\usepackage{bm}% bold math
\usepackage[utf8]{inputenc}
\usepackage[T1]{fontenc}
\usepackage{mathptmx}
\usepackage{etoolbox}
\usepackage{float}
\usepackage{amsmath}
\usepackage{xcolor}
\usepackage{siunitx}
\usepackage{mathrsfs}
\usepackage{ulem}

\author{Shadi Rezaei}
\affiliation{Dipartimento di Scienze Matematiche e Informatiche, Scienze Fisiche e Scienze della Terra, Università di Messina, Messina, Italy}
\alsoaffiliation[]
{Department of Physics, Faculty of Science, University of Kurdistan, Sanandaj, Iran}
\alsoaffiliation[]
{CNR-IPCF, Istituto per i Processi Chimico-Fisici, Messina, Italy}
\author{David Bronte Ciriza}
\affiliation{CNR-IPCF, Istituto per i Processi Chimico-Fisici, Messina, Italy}
\author{Abdollah Hassanzadeh}
\affiliation{Department of Physics, Faculty of Science, University of Kurdistan, Sanandaj, Iran}
\author{Fardin Kheirandish}
\affiliation{Department of Physics, Faculty of Science, University of Kurdistan, Sanandaj, Iran}
\author{Pietro G. Gucciardi}
\affiliation{CNR-IPCF, Istituto per i Processi Chimico-Fisici, Messina, Italy}
\author{Onofrio M. Maragò}
\affiliation{CNR-IPCF, Istituto per i Processi Chimico-Fisici, Messina, Italy}
\author{Rosalba Saija}
\affiliation{Dipartimento di Scienze Matematiche e Informatiche, Scienze Fisiche e Scienze della Terra, Università di Messina, Messina, Italy}
\email{rsaija@unime.it}
\phone{}
\fax{}
\alsoaffiliation[]
{CNR-IPCF, Istituto per i Processi Chimico-Fisici, Messina, Italy}
\author{Maria Antonia Iatì}
\affiliation{CNR-IPCF, Istituto per i Processi Chimico-Fisici, Messina, Italy}
\email{mariaantonia.iati@cnr.it}
\phone{}
\fax{}

\title[An \textsf{achemso} demo]
  {Faster calculations of optical trapping using neural networks trained by T-matrix data: an application to  micro and nanoplastics}

\keywords{optical trapping, T-matrix, machine learning, micro and nanoplastics}

\abbreviations{}

\begin{document}

%%%%%%%%
\begin{tocentry}

%\begin{figure*}[ht]
\centering\includegraphics[width=8.25cm]{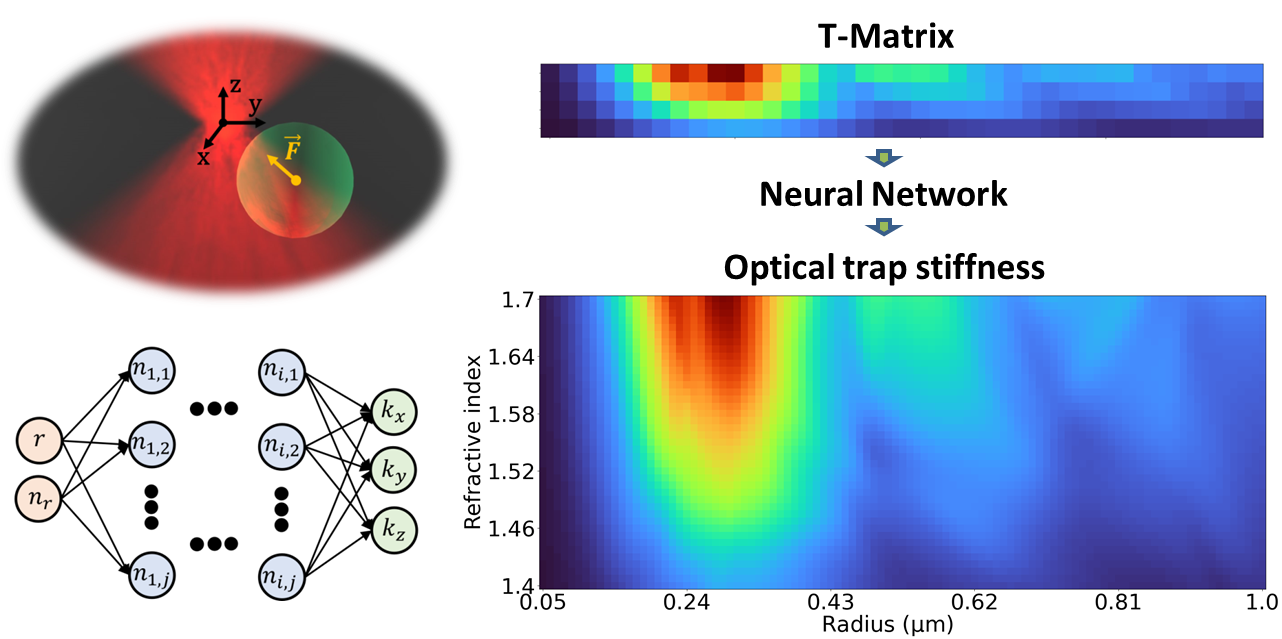}
%\label{Fig:TOC}
%\end{figure*}

%Some journals require a graphical entry for the Table of Contents.
%This should be laid out ``print ready'' so that the sizing of the

%Inside the \texttt{tocentry} environment, the font used is Helvetica
%8\,pt, as required by \emph{Journal of the American Chemical

%The surrounding frame is 9\,cm by 3.5\,cm, which is the maximum
%permitted for  \emph{Journal of the American Chemical Society}
%graphical table of content entries. The box will not resize if the
%content is too big: instead it will overflow the edge of the box.

%This box and the associated title will always be printed on a
%separate page at the end of the document.

\end{tocentry}

\begin{abstract}
We employ neural networks to improve and speed up optical force calculations for dielectric particles. The network is first trained on a limited set of data obtained through accurate light scattering calculations, based on the Transition matrix method, and then used to explore a wider range of particle dimensions, refractive indices, and excitation wavelengths.  This computational approach is very general and flexible. Here, we focus on its application in the context of micro and nanoplastics, a topic of growing interest in the last decade due to their widespread presence in the environment and potential impact on human health and the ecosystem.
\end{abstract}


\section*{Introduction}
Optical trapping\cite{jones2015optical,volpe2023roadmap} (OT) depends on size, material, and shape of the particles, as well as on some experimental parameters such as laser wavelength, numerical aperture (NA) of the objective lens, and quality of the focused beam. The design of optimal OT setups is a challenge that requires the support of extensive theoretical calculations of the trapping forces and stiffness to guide the selection of the best experimental parameters (power, wavelength, numerical aperture) to stably trap even the smallest particles\cite{marago2013optical,polimeno2018optical}. This can be time consuming and computationally demanding when carried out using the exact electromagnetic theory\cite{borghese2007}, especially when there is a need to investigate a wide range of models.

NNs have been shown to be an efficient approach to improve the speed of optical force calculation\cite{ciarlo2024deep} for spheres \cite{lenton2020machine} and more complex geometries, like ellipsoids \cite{bronte2022faster} or red blood cells\cite{tognato2023modelling}. NNs adjust their solutions to specific problems by training on data \cite{mitchell1997machine} and they have been used to improve the speed of conventional algorithms in a great variety of topics ranging from epidemics containment \cite{natali2021improving} to enhancing microscopy \cite{rivenson2017deep}, efficient tracking particles \cite{midtvedt2021quantitative}, and optical tweezers \cite{ciarlo2024deep}. Recently, NNs have also been used for the calculation of the scattering properties of spheroidal aerosols with prospects for applications to cosmic dust studies\cite{chen2022analytical,jing2023deep}. However, despite the great potential of NNs, only a few studies have been reported  about their use for the characterization of microplastics\cite{lorenzo2021deep,zhu2021microplastic, wegmayr2020instance,han2023deep}.

The widespread presence of plastic debris in terrestrial and aquatic environments has caused growing concern in the last decade and stimulated both political and scientific debate\cite{hale2020global}. A comprehensive study of microplastics (defined as plastic particles smaller than 5 mm \cite{frias2019microplastics, hale2020global}) and nanoplastics (smaller than $1\ \mu {\rm m}$ \cite{lambert2016characterisation,hartmann2019we}) is essential to assess their sources, distribution patterns, and potential implications, as well as to formulate effective strategies for environmental protection and human well-being.

Microplastics particles can be generated by manufacturing processes, clothing, makeup, food packaging, and industrial activities (primary microplastics) or can result from the fragmentation of larger particles  (secondary microplastics) due to environmental ageing. The generation of nanoplastics is a result of photodegradation or biodegradation of microplastics \cite{gigault2016marine, lambert2016characterisation, dawson2018turning}. 
The potential for transfer through the trophic chain \cite{farrell2013trophic} makes micro and nanoplastics particles (MNPs) a source of contamination at all trophic levels. This raises  toxicity concerns for animal and human health \cite{marfella2024microplastics}. The issue of risk assessment is very debated and challenging encompassing information about MNPs' occurrence, size distribution, morphology, physical, and chemical characteristics \cite{koelmans2022risk,le2023comprehensive}.

Size distribution of MNPs covers a wide range going from nanometers to millimetre size and has been shown to follow a power law with a negative exponent, depending on the formation process. This means that number concentrations increase with smaller sizes \cite{koelmans2022risk}. In spite of this evidence, the detection techniques for smaller particles still show strong limitations. 
Existing surveys have primarily focused on particles with sizes $\geq 20\ \mu {\rm m}$ \cite{cozar2014plastic,erni2017lost,suaria2016mediterranean,eo2018abundance}, with limited reports addressing the sub-$20\ \mu {\rm m}$ fraction \cite{pivokonsky2018occurrence,gigault2018current,xie2022strategies}. Although the use of optical tweezers (OT) for trapping microplastics is relatively novel, they appear as a promising technique for the trapping, manipulation, and characterization of MNPs thanks to their capability for a contactless investigation of single micro and nanoparticles\cite{marago2013optical,polimeno2018optical}. Recently, Raman tweezers\cite{petrov2007raman} (RT), where Raman spectroscopy is used to characterize the composition of a single particle trapped by optical tweezers\cite{ashkin1986observation,jones2015optical}, have emerged as a contactless, non-destructive, and uncontaminated tool to study small MNPs \cite{gillibert2019raman,ripken2021analysis,gillibert2022raman}. RT allow an analysis at the single particle level making it possible to detect and classify unambiguously MNPs, discriminating from other components and overcoming the capacities of standard Raman spectroscopy in liquid \cite{gillibert2019raman}.

In this work, we initially  model the optical trapping using a full electromagnetic description in the framework of the multipole field expansions and the Transition matrix (T-matrix) method \cite{borghese2007,waterman1971symmetry}. Then, the T-matrix data are used for the training of neural networks (NNs) that allow to extend the investigation of the optimal trapping configurations over a wide range of particle dimensions and refractive indices. We demonstrate the remarkable capability of NNs to accurately predict the optical trapping stiffness of MNPs on a broad set of cases, so addressing the experimental effort in the search of stable trapping conditions. Once the NNs have been trained, their use results in considerable savings in terms of computational time, with an acceptable loss in accuracy, thus paving the way to a faster identification of optimal trapping configurations for a wide variety of MNPs with different material composition and size.

\section*{Theory and modelling}
The goal of this work is to evaluate the strength of optical trapping on micro and nanoparticles of materials and dimensions that are of concern in microplastics and nanoplastics pollution research. We have considered target plastic materials such as Polyethylene (PE), Polyethylene terephtalate (PET), Polystyrene (PS), and Polyvynil alcohol (PVA), whose refractive indices are comprised between 1.4 and 1.7 (see Supplementary Table S1) \cite{gillibert2019raman}. The strength and efficiency of the trapping is evaluated by calculating the optical trap stiffnesses, $k_x$, $k_y$, $k_z$, defined as the slope of the linearized optical force components, $F_x$, $F_y$, $F_z$, around the equilibrium position, in all the three $x, y, z$ spatial directions\cite{polimeno2018optical}. For the sake of simplicity we report the stiffnesses normalized to the incident power, $P$. Calculations are carried out on spherical model particles dispersed in water of radius, $a$, from $50$ nm to $1$ $\mu$m.
Note that the sphere model is certainly an idealization, not representative of what we can actually find in the environment. However, the choice of using this simple model is due to the aim of this work which is to show the potential and limits of using NNs compared to the use of an accurate electromagnetic description.
We first calculate optical trapping forces at 785 nm, that is a common wavelength used in optical trapping and Raman spectroscopy experiments. In fact, near-infrared (NIR) excitation at 785 nm is generally favoured because of the absence of spurious fluorescence backgrounds due to contaminants in the water or the presence of eco-coronas around the particles. Then, we extend our investigation to other wavelengths generally used for optical trapping, 1064 nm,  and Raman spectroscopy, 515 nm. 

\begin{figure*}[th]
\centering\includegraphics[width=0.71\textwidth]{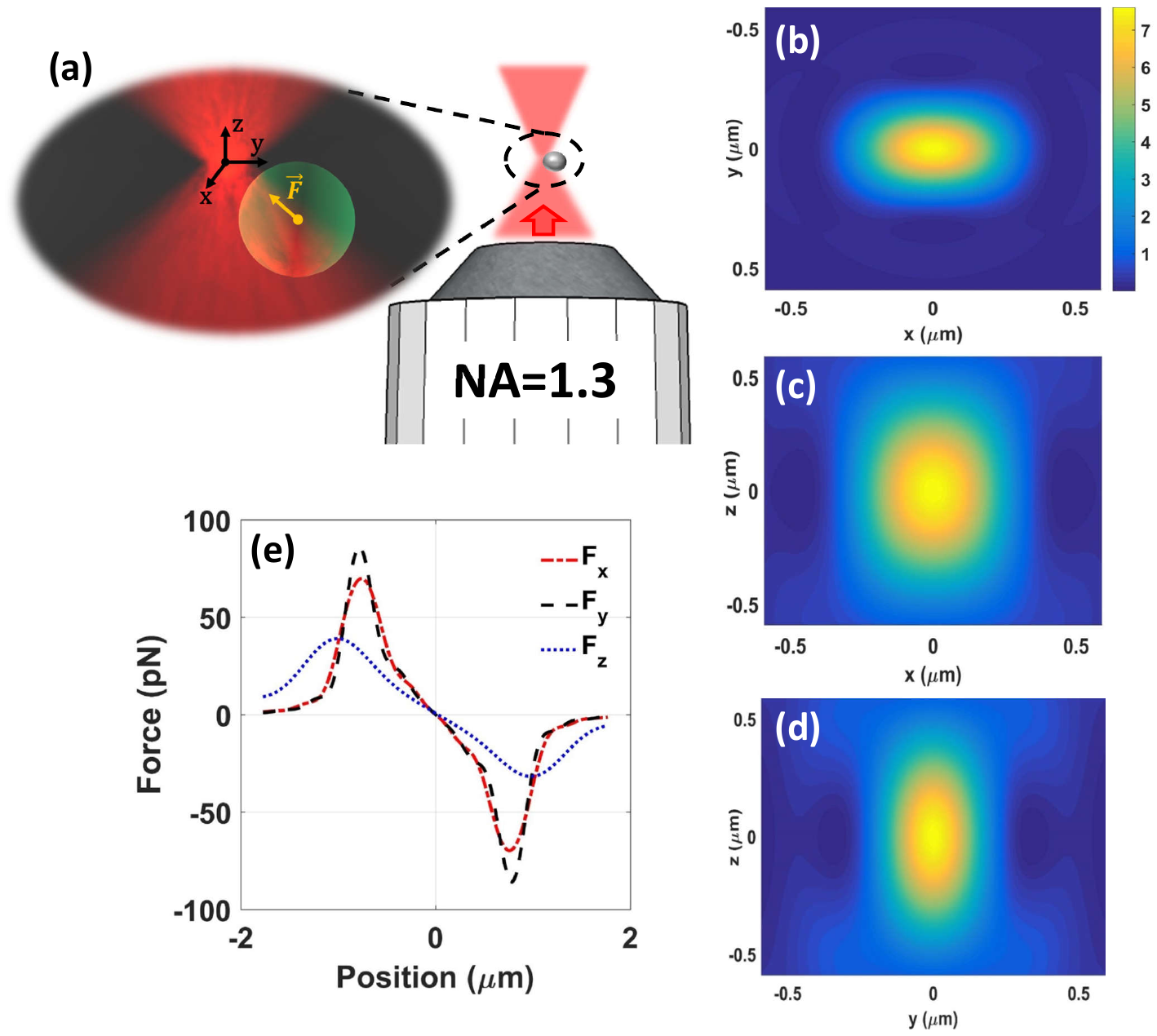}
\caption{(a) Sketch of the trapping geometry. A Gaussian TEM$_{00}$ $x-$polarized beam propagates along the axial $z-$direction and it is focused by a high NA objective lens. This produces a high intensity gradient in the lens focal region able to trap MNPs. (b-d) Maps of the focused field intensity normalized to the incident field, $|\frac{\mathit{E}}{\mathit{E_0}} |^2$ , in the $xy$ (b), $xz$ (c), and $yz$ (d) planes. We have considered incident light with wavelength $\lambda=785$ nm focused in water ($n_m = 1.33$) through a 1.3 $NA$ objective lens. The diffraction limited spot dimensions (FWHM) are $\Delta_x\approx 430$ nm, $\Delta_y\approx 285$ nm, and $\Delta_z\approx 615$ nm for the $x$, $y$, and $z$ direction, respectively. (e) Optical force components, $F_x$, $F_y$, $F_z$, as a function of the particle displacement from the nominal paraxial focus, calculated for a spherical particle with radius $a=0.8 \, \mu m$, refractive index $n_p=1.4$, immersed in water and illuminated by a laser beam focused as in (b-d).}
\label{Fig:1}
\end{figure*}
\paragraph{Strategy of the calculations.}
The workflow that we follow in the calculations is the following: we first model the system and calculate the main quantities involved in the trapping process using a full electromagnetic description in the framework of the multipole field expansion and the T-matrix method\cite{waterman1971symmetry,borghese2007}. Despite the fact that T-matrix provides accurate results, the use of this approach on a broad range of particle dimensions and composition  can be computationally demanding and time consuming. Thus, calculations are made on a limited set of sizes and materials (4 refractive index values, 39 particle size values for each refractive index). The results obtained using the T-matrix method are used for the training of NNs that allow us to accurately predict the normalized trap stiffnesses, $k_x/P$, $k_y/P$, $k_z/P$, for MNPs with radius and refractive index values never encountered during training thus saving computational time. Following this approach, we can obtain maps of the normalized trap stiffness for different wavelengths across a range of refractive indices and sizes typical of plastic materials. This can aid the identification of the optimal configurations for microplastics trapping significantly accelerating the exploration of optical trapping applications in MNPs research and classification.

\paragraph{Electromagnetic modelling of optical trapping in the T-matrix formalism.}
Optical forces and optical trapping are the result of the redistribution of linear momenta in a light scattering process \cite{jones2015optical}. In order to calculate optical trapping forces on MNPs we use electromagnetic scattering theory in the T-matrix formalism \cite{waterman1971symmetry,borghese2007} that provides a compact formalism based on the multipole expansion of the
fields. The geometry we model is shown in Fig. \ref{Fig:1}a, where a non-magnetic particle, with refractive index $n_p$, immersed in water (refractive index $n_m=1.33$) is illuminated by a Gaussian TEM$_{\rm 00}$ $x$-polarized laser beam propagating along the axial $z$-direction, tightly focused by a high numerical aperture ($NA=1.3$) objective lens producing a high intensity gradient in the lens focal region able to trap MNPs. 

The starting point of our calculations is the evaluation of the time-averaged radiation force 
 $\mathbf{F}_{\mathrm{rad}}$ through the integration of the Maxwell stress tensor\cite{jones2015optical}:

\begin{equation}\label{eq:radiation_force_and_torque}
\mathbf{F}_{\mathrm{rad}}= \oint_{S} \hat{\textbf{n}}\cdot \langle  \mathsf{T}_{\mathrm M}  \rangle
\ \mathrm{d}S \ , 
\end{equation}

\noindent where the integration is carried out over the surface $S$ surrounding the particle, $\hat{\textbf{n}}$ is the outward normal unit vector, $\textbf{r}$ is the vector
position, and $\langle \mathsf{T}_{\mathrm M} \rangle$ is the averaged Maxwell stress tensor in the Minkowski form which
describes the light-matter mechanical interaction\cite{pfeifer2007colloquium}. This can be simplified in terms of the incident and scattered field complex amplitudes since we consider always harmonic fields, at angular frequency $\omega$ in a homogeneous, linear, non-dispersive, and non-magnetic medium (water in our case) \cite{jones2015optical,borghese2007optical}, leading to:

\begin{equation}\label{eq:emt:eqC}
\mathbf{F}_{\rm rad} = -\frac{\varepsilon_{\rm m} r^2}{4} 
\int_{\Omega}
\left( |{\bf E}_{\rm s}|^2  + \frac{c^2}{n_{\rm m}^2} |{\bf B}_{\rm s}|^2 + 2  \mathrm{Re}\left\{ {\bf E}_{\rm i} \cdot {\bf E}_{\rm s}^{\ast} + \frac{c^2}{n_{\rm m}^2} {\bf B}_{\rm i} \cdot {\bf B}_{\rm s}^{\ast} \right\} \right)
\hat{\mathbf{r}} \, d\Omega \; ,
\end{equation}

\noindent where $\varepsilon_{\rm m}=\varepsilon_{\rm 0} n_m^2$ is the permittivity of the medium, $c$ is the speed of light in vacuum, $\hat{r}$ and $r$ are the unit vector and modulus of the vector position, the fields $\textbf{E}_{\mathrm i}$, $\textbf{B}_{\mathrm i}$, and $\textbf{E}_{\mathrm s}$,
 $\textbf{B}_{\mathrm s}$, are the incident electric and magnetic fields and the scattered electric and magnetic
fields, respectively, and the integration is over the full solid angle.
The key point of the T-matrix formalism is the expansion of the fields into a basis of vector spherical harmonics and
the consequent application of the boundary conditions across the particle surface\cite{borghese2007}. 
Thus, because of the linearity of the Maxwell equations, it is possible to define the T-matrix as the linear operator connecting the scattered fields to the incident ones. The T-matrix elements, 
$\{ T^{(p'p)}_{l'm'lm} \}$ with $(l,m)$ indices of the spherical harmonics and $p$ parity index, contain all the information on the morphology of the scatterer with respect to the incident fields, relating the known multipole amplitudes of the incident field $W^{(p)}_{{\mathrm i}, lm}$ with the unknown amplitudes of the scattered field $A_{{\mathrm s},l'm'}^{(p')}$ \cite{borghese2007}, $i.e.$,

\begin{equation}\label{eq:T_matrix_polarization}
A^{(p')}_{{\mathrm s}, l'm'} =  \sum_{plm} T^{(p'p)}_{l'm'lm} \; W^{(p)}_{{\mathrm i},lm}
.
\end{equation}

The special case of a homogeneous sphere (Mie theory) of radius $a$ results in a diagonal T-matrix,
independent of the index $m$, whose coefficients are related to the Mie coefficients, $a_l=-A_{{\mathrm
s},lm}^{(2)}/W_{{\mathrm i},lm}^{(2)}$ and $b_l=-A_{{\mathrm s},lm}^{(1)}/W_{{\mathrm i},lm}^{(1)}$
\cite{borghese2007}. We note that in the general case of inhomogeneous incident fields illuminating spherical particles the T-matrix formalism is equivalent to the generalised Lorenz-Mie theory\cite{gouesbet2017generalized,gouesbet2010t}.

For the case of optical tweezers, we consider an incident Gaussian beam focused by a high NA objective lens expressed in terms of its angular spectrum representation \cite{neves2006electromagnetic, borghese2007optical} and then calculate the optical forces and optical trapping stiffnesses for each axis of the trap\cite{jones2015optical}. In particular, we consider the multipole
amplitudes $\tilde{W}^{(p)}_{{\mathrm i}, lm}$ resulting from the field expansion around the focal point as:
\begin{equation}
\tilde{W}_{{\mathrm i},lm}^{(p)} ({\mathbf P})= \frac{ik_{\mathrm t}f e^{-ik_{\mathrm t}f} }{2\pi}
\int\limits_0^{\theta_{\mathrm{max}}} \sin\theta \int\limits_0^{2\pi} E_{\mathrm i}(\theta,\varphi)
\; W_{{\mathrm i},lm}^{(p)}(\hat{\mathbf k}_{\mathrm i},\hat{\mathbf e}_{\mathrm i}) \;
e^{i{\mathbf k}_{\mathrm t}\cdot{\mathbf P}} \ \mathrm{d}\varphi \ \mathrm{d}\theta  ,
\end{equation}

\noindent in which $f$ is the focal length, $\theta_{\mathrm{max}}=\arcsin(\mathrm{NA}/n_m)$,
$k_{\mathrm t}$ is the wavenumber transmitted through
the objective lens, and each transmitted plane wave has been %$\mathbf{E}_{\mathrm {ff},t}(\theta,\varphi)$,
expanded into multipoles\cite{jones2015optical} with amplitudes $W_{{\mathrm i},lm}^{(p)}(\hat{\mathbf k}_{\mathrm i},\hat{\mathbf e}_{\mathrm i})$. The centre around which the expansion is performed is considered displaced by
${\mathbf P}$ with respect to the focal point ${\mathbf O}$ and the amplitudes
$\tilde{W}^{(p)}_{\mathrm{i},lm}({\mathbf P})$ define the focal fields and can be numerically
calculated by knowing the characteristics of the optical system. The expression for the
radiation force along the direction of a unit vector $\hat{\mathbf u}$, $i.e.$, $F_{\mathrm
{rad}}(\hat{\mathbf u}) = {\mathbf F}_{\mathrm {rad}} \cdot \hat{\mathbf u}$ can be obtained
through the knowledge of the scattered amplitudes $\tilde{A}^{(p)}_{{\mathrm s},lm}$ related to the
incident focal fields through the particle T-matrix\cite{polimeno2018optical}.

We note that to ensure the numerical stability of the results and an accurate representation of the fields and of all the other observables (cross sections, optical forces, stiffnesses), the sums related to the multipole expansions must be
extended to a sufficiently high value of the index $l$, called $l_{\rm M}$, 
 so that when the truncation index $l_{\rm T}>l_{\rm M}$ the computational results for the chosen quantity are stable within the desidered accuracy \cite{saija2003efficient}. This truncation value $l_{\rm M}$ depends in general on the size, shape, and composition of the particles.

\paragraph{Machine learning for optical trapping.}
The calculation of optical forces often faces a trade-off between speed and accuracy. Machine learning allows to overcome this limitation, achieving fast and accurate results \cite{lenton2020machine, bronte2022faster}. The use of NNs is expanding beyond optical force calculations, enhancing the calibration of optical tweezers \cite{argun2020enhanced}, improving particle tracking\cite{helgadottir2019digital}, and optimizing the design of optical tweezers setups\cite{li2019algorithmic}. Without the use of NNs, these tasks require manual tuning of parameters, low noise measurements, or extremely long calculations \cite{ciarlo2024deep}.  

Here we selected a neural network architecture with 2 layers of densely connected neurons as it represents the simplest model capable of achieving our desired results. For each neuron in our NN we utilized the sigmoid activation function. The sigmoid function was chosen for its smooth gradient, which aids in the training process by providing clear gradient signals for backpropagation. 
This NN choice was driven by the need for efficiency in our specific context of force calculation in optical tweezers. In fact, while convolutional NNs are quite popular, their primary strength lies in image analysis due to their ability to capture spatial hierarchies through convolutional layers \cite{mitchell1997machine}. In contrast, force calculation in optical tweezers involves processing numerical data where densely connected layers are typically more suitable. Furthermore, densely connected neural networks have been commonly employed in similar force calculation tasks within optical tweezers applications, proving their effectiveness and reliability in this context \cite{ciarlo2024deep,lenton2020machine,bronte2022faster,tognato2023modelling}. Therefore, we opted for a dense architecture to align with the established methods and ensure optimal performance for our specific application.

The NNs training procedure involves several steps. First, the architecture definition consists of choosing the number of layers and the number of neurons per layer. The complexity of the problem under study drives the choice of the architecture: a higher number of trainable parameters produces a model capable of learning more from the training data. Second, pre-processing the data and dividing them into a training, validating, and test data set is required.  The iterative part of the training starts by loading a subset of the training data and applying the training step where the NNs weights are optimized to minimize the loss function. We use the mean squared error as the loss function and the Keras implementation of the Adam optimizer with the default parameters \cite{chollet2018keras}.  Once the training dataset has been fully explored, the error between the NNs calculation and the validating dataset (defined as the mean square difference) is computed. The iterative step is repeated until this error stops decreasing and the final error is evaluated against the test data set.

\section*{Results and discussion}
\paragraph{T-matrix results.}
Here we show the results of our computations of the optical trap stiffnesses of model MNPs in water using the T-matrix approach.

\begin{figure*}[ht]
\centering\includegraphics[width=1\textwidth]{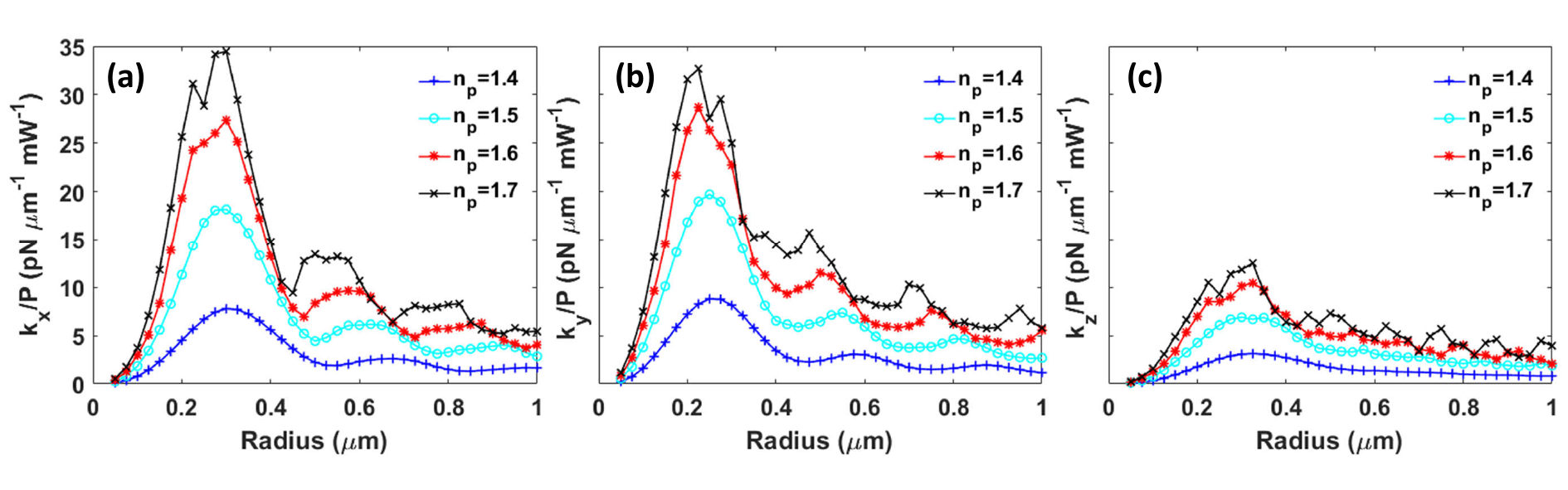}
\caption{Trap stiffnesses per unit  laser power $k_x/P$ (a), $k_y/P$ (b), $k_z/P$ (c), calculated for spherical particles with refractive index $n_p$=1.4, 1.5, 1.6 and 1.7 (see legend for the different colors), as a function of particle radius in the range [0.05 $\mu m$  - 1$\mu m$]. The particles are immersed in water ($n_m = 1.33$), $\lambda=785$ nm, and  NA$=1.3$.
}
\label{Fig:2}
\end{figure*}

We have first calculated the electromagnetic field intensity distribution at 785 nm as it is focused by a high numerical aperture objective (NA = 1.3) in water ($n_m = 1.33$), through the angular spectrum representation \cite{neves2006electromagnetic,borghese2007optical}. Figures \ref{Fig:1} (b-d) show the maps of the calculated focused field intensity normalized to the incident field intensity, $|\frac{\mathit{E}}{\mathit{E_0}} |^2$ , in the $xy$ (b), $xz$ (c), and $yz$ (d) planes.  In the $xy$ plane the cylindrical symmetry of the intensity map is broken by the field polarization, resulting in a tighter spot size along $y$ (FWHM of $\Delta_y \approx 285$ nm) with respect to the $x$ polarization direction ($\Delta_x \approx 430$ nm), while in the $xz$ and $yz$ planes the field appears elongated due to beam propagation ($\Delta_z \approx 615$ nm). Thus, we calculate the optical forces acting on the particles  as a function of particle size and composition. To calculate the transverse force on the particle at the equilibrium position, the $z$ (longitudinal) coordinate at which the axial force vanishes is found. Thus, the force components in the transverse directions ($x$, $y$) are calculated. As an example, in panel (e) we plot the  optical force components, $F_x$, $F_y$, $F_z$, as a function of the particle displacement from the nominal paraxial focus, for a sphere with radius $a=0.8 \, \mu m$, refractive index $n_p=1.4$, illuminated by a  laser beam focused as in (b-d)  with power 30 mW. We note that the force component  $F_z$ is smaller than the two other in-plane components and that the trapping position of the particle in the  $z$-direction, $F_z(z_0)=0$, is typically offset from the centre of the coordinate system due to the offset induced by the optical scattering force. Finally, we observe that for the considered laser power, optical forces of the order of several tens of pN are obtained.

In proximity of the equilibrium point the optical force can be linearized as an elastic restoring force with negative slope, \textit{e.g.}, $F_x\approx-k_xx$ for the $x$-direction. Within this linear range, optical tweezers are approximated by an effective harmonic potential with trap stiffnesses $k_x$, $k_y$, $k_z$. In order to calculate the optical trap stiffnesses, we get the slope of the force-displacement graphs at the equilibrium position, where the force vanishes \cite{polimeno2018optical}. 

In Fig.\ref{Fig:2} we show the calculated stiffness normalized to laser power $k_x/P$ (a), $k_y/P$ (b), $k_z/P$ (c)  as a function of the particle radius, in the [0.05 $\mu m$  - 1$\mu m$] range. We model the particle composition assuming four different values of refractive index $n_p=1.4, 1.5, 1.6, 1.7$, chosen in a range representative of plastics (see supplementary Table S1)\cite{gillibert2019raman}. In the three panels in Fig.\ref{Fig:2}, for all the assumed refractive index values, we observe maximum values  around $5-35$ pN $\mu$m$^{-1}$ mW$^{-1}$ corresponding to a particle radius  $a\approx 0.25-0.35 \, \mu m$. This is the size range for which the volume of the particle overlaps with the laser spot, optimizing the interaction region\cite{neto2000theory,polimeno2018optical}. The peak in the axial $z-$direction appears lower compared to those in the $x-$, $y-$ transverse directions. The explanation for this behaviour appears clear from the shape of the focal spot in Fig.\ref{Fig:1}(b-d) which is less tightly focused in the  $z-$direction. Thus, the longitudinal trap stiffness, $k_z$, gives an evaluation of how strongly a particle can be trapped in 3D \cite{knoner2006calculation}. Note that even when axial trapping is not occurring, transverse optical forces can always grant guiding, channeling, and aggregation of particles also at low power and at the nanoscale\cite{bernatova2019wavelength}. For small particle size the trap stiffness has a volumetric scaling related to the scaling of the induced dipole interaction between particle and fields\cite{polimeno2018optical}.
As the particles become larger than the interaction region, we can observe a decrease of the stiffness and the onset of a modulation in the curve due to the interference between different multipoles. The dependence of the trap stiffness on particle size and refractive index is not linear but shows maxima and minima, corresponding to strengthening and weakening of the trapping force. This is due to constructive or destructive interference typical of the interference structure of Mie scattering\cite{stilgoe2008effect}.

From a quantitative point of view, the calculated trapping stiffnesses $k_x$, $k_y$, $k_z$ on a particle of radius $a\approx 0.25-0.35 \, \mu$m and refractive index $n_p = 1.7$ are larger by a factor $\approx 7$ (for $k_x$, $k_y$) and $\approx 4$ (for $k_z$) with respect to the case of the trapping of a particle of radius  $a\approx 1 \, \mu$m. The trend is the same, although with slightly reduced ratios, also for particles of lower refractive index. The higher stiffness values indeed improve the trapping stability for smaller nanoparticles which are more subject to Brownian motion than larger microparticles\cite{volpe2013simulation}. In fact, due to the increased size, thermal fluctuations have a smaller influence on the trapping stability of micron scale particles, permitting stable 3D trapping even with low trap stiffnesses. For what concerns nanoparticles with radius lower than  $0.25\, \mu$m, it is evident how the trapping stiffness scales with the particle volume ($\approx a^3$), reaching down values of $0.05-0.5$ pN $\mu$m$^{-1}$ mW$^{-1}$ for particles of radius $a =50$ nm. This is due to the volumetric scaling of the particle polarizability that controls optical forces at the nanoscale\cite{marago2013optical}. In experiments thermal fluctuations contribute to a destabilization of the optical trapping process. Thus, increasing the laser power is required to achieve stable 3D trapping.  

The investigation of the dependence of the trapping stiffness on the particle dimensions has been extended to particle radii up to 2.5 $\mu$m (see Supplementary Note S1) and for different wavelengths. For this reason,  to ensure the convergence of the multipole field expansions over the entire parameter range, according to the Wiscombe criterion \cite{wiscombe1980improved}, we adopted a truncation index $l_{\rm M}$=20 throughout the computations.

\begin{figure*}[ht]
\centering\includegraphics[width=1\textwidth]{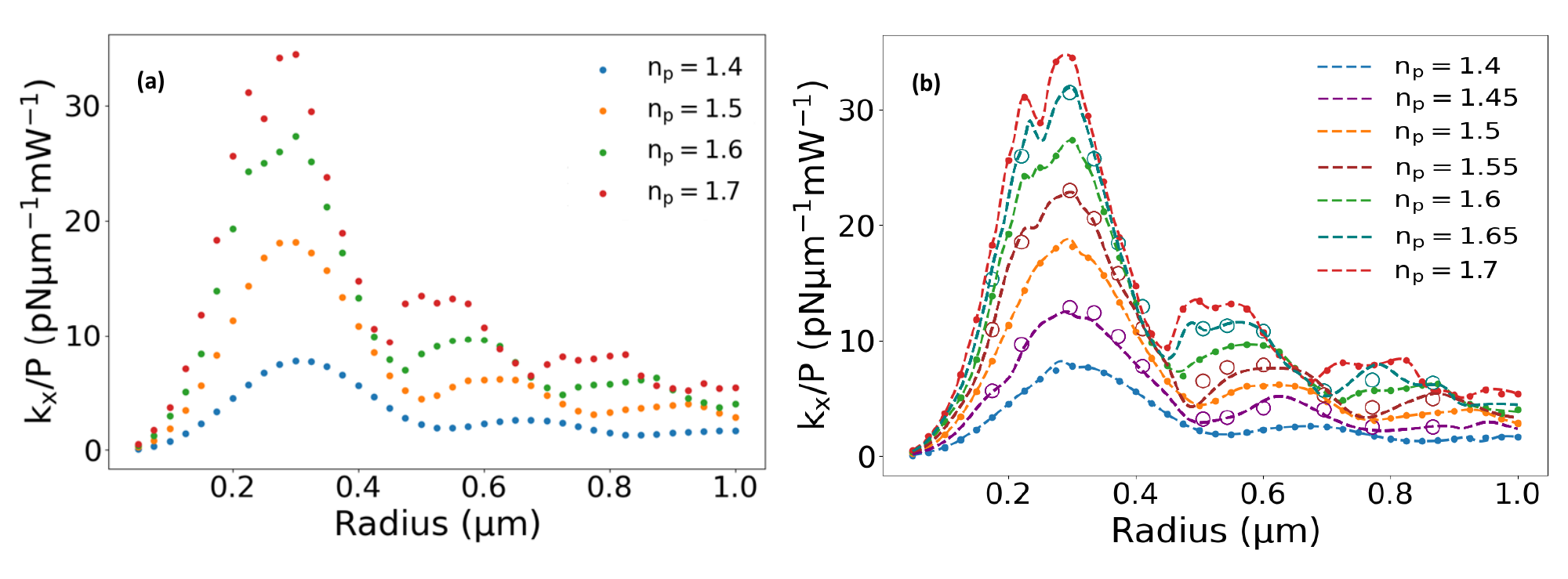}
\caption{Transverse stiffness normalized to laser power $k_x/P$ as a function of radius at a laser wavelength of 785 nm and with NA=1.3. (a) Dots represent the 156 values computed using the T-matrix technique for $n_p$ = 1.4, $n_p$=1.5, $n_p$=1.6, and $n_p$=1.7 and are the dataset used for training the NN. (b) Dashed lines represent the NN predictions for different refractive indices. Dots represent the exact T-matrix stiffness values used for the NN training (as in panel a). The open circles represent the T-matrix values computed for the new set of refractive indices  $n_p$ = 1.45, $n_p$=1.55, $n_p$=1.65, that we use as sampling points to check the validity of the NN predictions.} 
\label{Fig:3}
\end{figure*}

\paragraph{Machine learning results.}
Despite the high level of accuracy of the results obtained using the T-matrix method, this approach can be computationally demanding and time consuming. The dataset plotted in Fig.\ref{Fig:2} is relatively small, including 156 data points, i.e. 39 points for each of the four refractive indices $n_p=1.4, 1.5, 1.6, 1.7$, distributed in the particle radius range [0.05 $\mu m$  - 1$\mu m$]. Nevertheless, it required about  4 hours to be gathered. The same dataset was used to train the NNs, as shown in Fig. \ref{Fig:3}.  The T-matrix data for $k_x/P$ used for the training are reported in Fig. \ref{Fig:3}a where dots represent the 156 overall computed data, for the same radius and refractive index values as in Fig. \ref{Fig:2}a. These results represent all the available $k_x/P$ data used for the training of the NN at the specific wavelength $\lambda=785$ nm and for NA=1.3. NN calculations were subsequently done in order to predict the trap stiffness on a much larger set of data points,  considering different plastic materials with refractive indices $n_p$ = 1.45, $n_p$=1.55, and $n_p$=1.65 (see \ref{Fig:3} b), including 501 data points for each curve, with an overall computational time of less than 20 minutes, mostly due to the time needed for the training procedure. The training needs to be performed only once. After that, the NNs can provide results in computational times of the order of seconds or fractions of a second.

The NNs are modelled and trained in Python using Keras (version 3.11.4-tf) \cite{chollet2018keras} with TensorFlow backend (version 2.14.0). The NN training and T-matrix calculations were performed using an Intel Core i9- 10900X processor with 128 GB of RAM.

\begin{figure*}[ht]
\centering\includegraphics[width=1\textwidth]{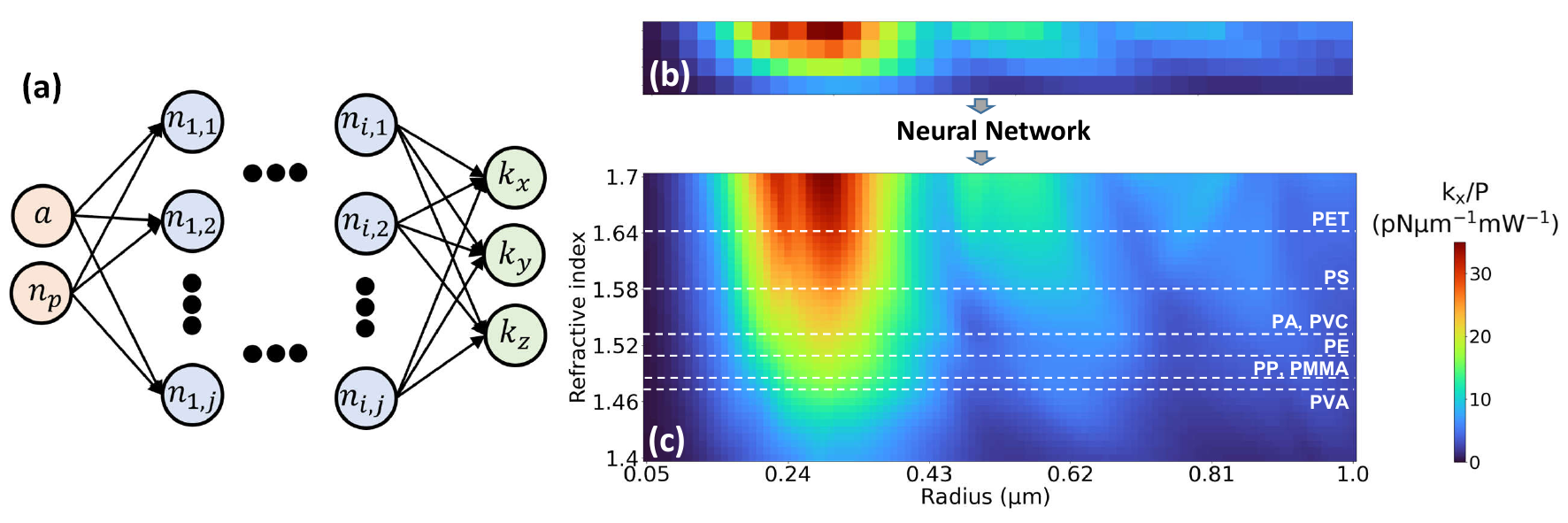}
\caption{(a) Architecture of the densely connected NNs used for the calculations with an input layer (pink), an output layer (green), and \textit{i} hidden layers (light blue). (b,c) 2D density plots of $k_x/P$ as a function of radius and refractive index. The 2D density plot in (b) is obtained plotting the 156 training points computed with the T-matrix, the density plot in (c) is generated with the NNs. The refractive indices of the most common plastic materials are shown as white dashed lines (PET, PS, PA, PVC, PE, PP, PMMA, and PVA).}
\label{Fig:4}
\end{figure*}

The architecture of the densely connected NNs used for the calculations is shown in Fig.\ref{Fig:4}(a) with an input layer (pink), an output layer (green), and three \textit{i} hidden layers (light blue). We have 2 degrees of freedom: the particle radius, $a$, and the particle refractive index, $n_p$. Each of the hidden layers has $\textit{j}=32$ neurons ($\approx 2.3\cdot 10^3$ trainable parameters) and all the neurons in each layer are connected to all the neurons in the previous and the next layer. 

Despite the limited training dataset, the NNs are able to effectively generalize and successfully predict trap stiffness values for new refractive indices, never encountered during the training. In Fig.\ref{Fig:3}(b) the NN predictions (dashed lines) are superimposed to the T-matrix data (full dots), used for the training procedure of the NN, for $n_p=1.4, 1.5, 1.6, 1.7$. Fig.\ref{Fig:3}(b) also shows the normalized stiffness for the new set of materials featuring $n_p$ = 1.45, 1.55, and 1.65 (dashed lines). In order to check the accuracy of the NN predictions, we performed exact T-matrix calculations of $k_x/P$ for the new materials with $n_p$ = 1.45, 1.55, and 1.65 for $N=36$ points and plotted the results (open circles). On these values we can estimate the accuracy by considering the discrepancy, $\delta$, between the NN predictions, $k_i^{\rm NN}$, and the T-matrix calculations, $k_i^{\rm TM}$:
\begin{equation}
\delta=\frac{1}{N}\sum_{i=1}^N \frac{|k_i^{\rm NN}-k_i^{\rm TM}|}{k_i^{\rm TM}}.
\end{equation}
For the data shown in Fig. \ref{Fig:3}b we obtain a discrepancy of $\delta_x\sim 0.065$, while in the $y$ and $z$ directions we obtain $\delta_y\sim 0.035$ and $\delta_z\sim 0.089$. However, the loss in accuracy of the NN results compared with the T-matrix ones is balanced by the significant savings in terms of computational times, as reported above. These results prove the capability of NNs to extend reliable predictions to intermediate particle size, interpolating complex relationships in a much shorter time. Furthermore, we can compare the accuracy of NN against a spline interpolation (see Supplementary Note S2) by considering the same sampling points at different radii as in Fig. \ref{Fig:3}b. The discrepancy for the spline interpolated data with respect to T-matrix data shows similar value as the one calculated for the NNs data. However, the spline interpolation cannot be successfully used to extrapolate away from the interpolation points yielding a discrepancy several times larger than the NN one (see Supplementary Fig. S3).

In order to take full advantage of the NN capabilities we have extended the calculations of $k_x/P$ to a much denser set of refractive indices and particle size values. To better display the results we use 2D density plots, in which we map the value of $k_x/P$ in color tones (increasing from blue to red), as a function of the particle radius (x-axis) and of the refractive index (y-axis). In Fig.\ref{Fig:4} we compare the 2D density plot (b), obtained plotting the 156 points from the T-matrix calculations, with  the much more detailed plot (c) generated using the NN, containing over 4000 points. Dashed horizontal white lines are added in correspondence of the refractive index values of the most common plastic materials such as PET (polyethylene terephthalate), PS (polystyrene), PA (polyamide), PVC (polyvinyl chloride), PE (polyethylene), PP (polypropylene), PMMA (polymethyl methacrylate), and PVA (polyvinyl alcohol). For all plastics (dashed white lines) the highest trapping stiffness, $k_x/P\approx15-30$ pN$\mu$m$^{-1}$m$W^{-1}$, is expected for particles with radius in the 200-400 nm range, i.e., in the nanoplastics regime. For larger particles (up to $a=1\mu$m) the trap stiffness slowly reduces down to 2.5-5 pN$\mu$m$^{-1}$m$W^{-1}$. For nanoplastics of radius $a\approx100$ nm  we have similarly stiffness of the order of few piconewton. However, the scaling is volumetric ($\approx a^3$).
The comparison between T-matrix and NN computational times is striking. Once the training procedure has been done, the density plot in Fig. \ref{Fig:4}c is generated in a fraction of a second using the NN and contains over 4000 points.  T-matrix computations would require more than 4 days to  generate a density plot with the same number of points. In this case the use of  NN reduces the computational time by 5 orders of magnitude.

\begin{figure*}[ht]
\centering\includegraphics[width=1\textwidth]{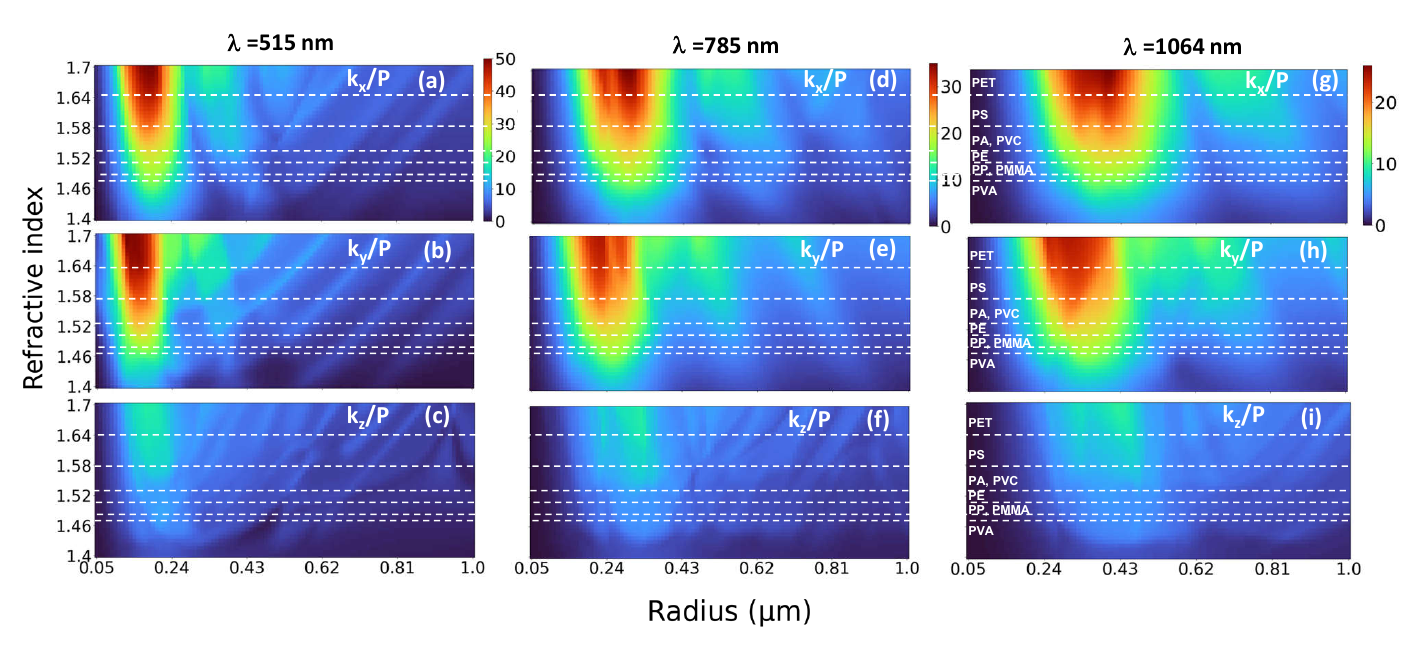}
\caption{2D density plots of stiffnesses normalized to laser power along the transverse ($x-y$) and longitudinal ($z$) directions, as a function of radius and refractive index at three wavelengths of the incident laser beam, $\lambda=515$ nm (a,b,c), $\lambda=785$ nm (d,e,f) and $\lambda=1064$ nm (g,h,i). The plots are obtained using the NNs predictions. White dashed lines are representative of the most common plastic materials as in Fig. \ref{Fig:4}c. Note the different scale bar for each wavelength.}
\label{Fig:5}
\end{figure*}

In order to look for an optimal trapping configuration, we extend our investigation to other typical wavelengths of the incident  laser beam used in optical tweezers experiments. In Fig. \ref{Fig:5} we show the 2D density plots of stiffnesses normalized to laser power along the transverse ($x-y$) and longitudinal ($z$) directions, as a function of radius and refractive index at three wavelengths, $\lambda=515$ nm (a,b,c), $\lambda=785$ nm (d,e,f) and $\lambda=1064$ nm (g,h,i). The plots are produced using the NNs.
It is evident from the figure (note the different scale bars) that the highest values of trapping stiffness is obtained for the shortest wavelength, $\lambda=515$ nm, and that trapping stiffness decreases at longer wavelengths.  In fact, the size of the diffraction limited spot, proportional to $\lambda$, increases with longer wavelengths. Thus, for the same incident power the intensity gradient reduces and hence the trapping force is also reduced.  Moreover, as expected, as the laser wavelength increases, the maximum trapping stiffness occurs for larger radii related to the increase of the spot size. 

\section*{Conclusions}
In summary, we have proposed a computational approach, based on the use of NNs trained with data obtained by accurate electromagnetic calculations based on the T-matrix method, capable to predict the best optical trapping configurations of MNPs in water. We have investigated the variation of optical forces and optical trapping as a function of size and refractive index of MNPs, also considering different wavelenghts for the incident laser beam.
While training a NN can be time-consuming, the resulting advantages are substantial. The lengthiest part of the process is generating the training data, but this can be accelerated by parallelizing the calculations. Once the NN is trained, two primary benefits are realized. Firstly, the increase in computational speed  enables the exploration of scenarios that were previously beyond the reach of the electromagnetic calculations. Secondly, a trained NN is more straightforward to utilize and integrate with other software compared to existing methods\cite{bronte2022faster}.

The remarkable capability of NNs to generalize the calculations offers a broader perspective providing new insights and challenges into the optical trapping of MNPs. This approach also paves the way for the inclusion of additional parameters, considering  more complex shapes for the MNPs, for instance plastic nanofibers from textile and fabrics could be modelled as micro-chains of spherical subunits\cite{polimeno2021t}. On the other hand, this type of computational approach is extremely general and flexible and open perspectives for many other applications from the nanoscale\cite{marago2013optical} to space applications\cite{polimeno2021optical,magazzu2022investigation}.

\begin{acknowledgement}
This work has been funded by European Union (NextGeneration EU), through the MUR PNRR project SAMOTHRACE (ECS00000022), MUR PNRR project PE0000023-NQSTI, and PRIN2022 "EnantioSelex (2022P9F79R), “SEMPER” (20227ZXT4Z), “FLASH-2D” (2022FWB2HE), “PLASTACS” (202293AX2L)", Exo-CASH" (2022J7ZFRA), and "Cosmic Dust II" (2022S5A2N7).
\end{acknowledgement}

\begin{suppinfo}
The Supporting Information is available free of charge on the
ACS Publications website. Table S1: Refractive index of common plastic materials. Note S1: Extension to larger particles; Note S2: Comparison with spline interpolation. Figure S1: Calculations of trap stiffnesses for a size range up to $2.5\ \mu$m; Figure S2: 2D density plots of trap stiffnesses for a size range up to $2.5\ \mu$m; Figure S3: Discrepancy between NN and T-matrix compared to the one for spline interpolated data and T-matrix, as a function of the refractive index.
\end{suppinfo}

%\bibliography{acs-achemso}

\providecommand{\latin}[1]{#1}
\makeatletter
\providecommand{\doi}
  {\begingroup\let\do\@makeother\dospecials
  \catcode`\{=1 \catcode`\}=2 \doi@aux}
\providecommand{\doi@aux}[1]{\endgroup\texttt{#1}}
\makeatother
\providecommand*\mcitethebibliography{\thebibliography}
\csname @ifundefined\endcsname{endmcitethebibliography}
  {\let\endmcitethebibliography\endthebibliography}{}

%\newpage

%\begin{figure*}[ht]
%\centering\includegraphics[width=8.25cm]{TOC.png}
%\label{Fig:TOC}
%\end{figure*}

\end{document}